\documentclass[10pt,brazil,english]{article}
\usepackage[T1]{fontenc}
\usepackage[latin1]{inputenc}
\usepackage{a4wide}
\usepackage{color}
\usepackage{graphicx}

\makeatletter

\usepackage{babel}
\makeatother
\begin{document}

\title{Realism in Energy Transition Processes: an example from Bohmian Quantum
Mechanics}

\author{\textbf{J. Acacio de Barros, J. P. R. F. de Mendonça}\\
Departamento de Física -- ICE \\
Universidade Federal de Juiz de Fora\\
36036-330, Juiz de Fora, MG, Brazil\\
\\
\textbf{N. Pinto-Neto}\\
Centro Brasileiro de Pesquisas Físicas\\
R. Dr. Xavier Sigaud 150\\
22290-180, Rio de Janeiro, RJ, Brazil}

\maketitle
\begin{abstract}
In this paper we study in details a system of two weakly coupled harmonic
oscillators. This system may be viewed as a simple model for the interaction
between a photon and a photodetector. We obtain exact solutions for
the general case. We then compute approximate solutions for the case
of a single photon (where one oscillator is initially in its first
excited state) reaching a photodetector in its ground state (the other
oscillator). The approximate solutions represent the state of both
the photon and the photodetector after the interaction, which is not
an eigenstate of the individual hamiltonians for each particle, and
therefore the energies for each particle do not exist in the Copenhagen
interpretation of Quantum Mechanics. We use the approximate solutions
that we obtained to compute bohmian trajectories and to study the
energy transfer between the two particles. We conclude that even using
the bohmian view the energy of each individual particle is not well
defined, as the nonlocal quantum potential is not negligible even
after the coupling is turned off. 
\end{abstract}

\section{Introduction}

The discussions about the incompleteness of the wavefunction to describe
physical processes dates back to the beginning of quantum mechanics
itself. This discussion is closely related to the possibility of describing
quantum mechanical systems from an underlying realistic model. In
1952, David Bohm showed that such a realistic model was possible.
However, Bohm's theory had the problem of being nonlocal \cite{Bohm1,Bohm2}.
In 1963 John Bell showed that in order to obtain the same results
predicted by quantum mechanics, any realistic theory would have to
be nonlocal \cite{SuppesRepresentation}. Bell's result and the failure
of using the Copenhagen interpretation of quantum mechanics to some
particular situations, as in for example Quantum Cosmology, lead to
a raised interest in Bohm's interpretation and in nonlocal realistic
theories \cite{QuantumCosmology}. 

The subject of reallity and nonlocality has been an interest of Patrick
Suppes for quite a while \cite{SuppesRepresentation}, in particular
for the photon. In fact, one of the authors of this paper co-published
with him a series of papers that layed down the foundational analysis
of realistic and local model of photons that could explain the double
slit experiment, the EPR experiment and other phenomena \cite{PatAcacio1,PatAcacio2,PatAcacio3,PatAcacio4}.
The problem with the Suppes and de Barros model was that it did not
have a consistent theory of photon-counting for single photons, and
therefore could not explain the non-locality of single photons and
the GHZ experiment, for example. 

In this paper we try to respond, within Bohm's model, the question:
what is a photon? We do not follow the standard Bohmian interpretation
for bosonic fields (as can be found in \cite{Holland}). Instead,
we use the simple interpretation that {}``a photon is what a photodetector
detects''. One may think of a photodetection as a transfer of energy
from a quantized mode of the electromagnetic field (the photon) to
an atom in its ground state (the quantum photodetector). Therefore,
to study this photodetection we will focus on the process of transfer
of energy from the photon to the photodetector. 

To study the exchange of energy in details, we have to choose between
two different and simple models of a photo-detector: a photo-detector
with discrete or continuous band \cite{Cohen_Atom_Photon}. For the
purpose of simplicity, we will choose the former. However, since we
are only interested in the aspects of energy transfer between the
two systems, we will make an even further simplification and consider
that the photon and the detector will both be described by a single
harmonic oscillator. Furthermore, during some time $\Delta T_{int}$,
we will assume that a linear interaction exists between the two oscillators,
and that this interaction is weak. This detection model is known as
an indirect measurement \cite{QuantumMeasurement}, and has been the
subject of intense research lately as it is directly connected to
quantum nondemolition experiments. As we will see, this {}``toy model''
will allow us to capture some important features of the entanglement
between the two systems. 

This paper is organized in the following way. In Section \ref{sec:The-Classical-Case}
we will quickly review the interaction between two harmonic oscillators
for the classical case. This will allow us to understand how the transfer
of energy happens in such case. We then compute the exact solutions
for the quantum mechanical system with interaction (Section \ref{sec:Quantum-Evolution:-Exact}).
In Section \ref{sec:A-Simple-Example} we apply the results of Section
\ref{sec:Quantum-Evolution:-Exact} to a specific case of exchange
of a single quantum of energy and analyze its outcomes. In Section
\ref{sec:Bohm} we use Bohm's theory to interpret the results obtained.
The conclusions are in Section \ref{sec:Conclusions}.

\section{The Classical Case\label{sec:The-Classical-Case}}

\begin{figure}[htbp]
\begin{center}\includegraphics[%
  scale=0.3]{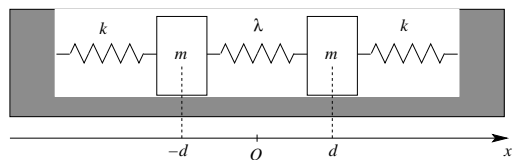}\end{center}

\caption{\label{Figure:Identical-harmonic-oscillators}Identical harmonic
oscillators coupled by a spring of constant $\lambda$. }
\end{figure}

Before we go into the details of the quantum mechanical examples,
let us begin by analyzing the classical system of two one-dimensional
coupled harmonic oscillators with the same mass $m$, elastic constant
$k$, and coupling constant $\lambda$, as shown in Figure \ref{Figure:Identical-harmonic-oscillators}.
The Hamiltonian for this system is given by 

\begin{equation}
H=\frac{1}{2m}\left(P_{1}^{2}+P_{2}^{2}\right)+\frac{1}{2}k\left(\left(X_{1}+d\right)^{2}+\left(X_{2}-d\right)^{2}\right)+\frac{1}{2}\lambda\left(X_{1}-X_{2}+2d\right)^{2}.\label{Classical_Hamiltonian_0}\end{equation}
 To simplify the equations of motion and eliminate the undesirable
constant $d$ we can make the canonical transformation \begin{eqnarray*}
x_{1} & = & X_{1}+d,\\
x_{2} & = & X_{2}-d,\\
p_{1} & = & P_{1},\\
p_{2} & = & P_{2}.\end{eqnarray*}
With the new variables equation (\ref{Classical_Hamiltonian_0}) rewrites
to\begin{equation}
H=\frac{1}{2m}\left(p_{1}^{2}+p_{2}^{2}\right)+\frac{1}{2}k\left(x_{1}^{2}+x_{2}^{2}\right)+\frac{1}{2}\lambda\left(x_{1}-x_{2}\right)^{2}.\label{Classical_Hamiltonian}\end{equation}
The Hamiltonian equations of motion are

\begin{eqnarray*}
\dot{p}_{1} & = & -\frac{\partial H}{\partial x_{1}}=-kx_{1}-\lambda\left(x_{1}-x_{2}\right),\\
\dot{x}_{1} & = & \frac{\partial H}{\partial p_{1}}=\frac{p_{1}}{m},\end{eqnarray*}
\begin{eqnarray*}
\dot{p}_{2} & = & -\frac{\partial H}{\partial x_{2}}=-kx_{2}+\lambda\left(x_{1}-x_{2}\right),\\
\dot{x}_{2} & = & \frac{\partial H}{\partial p_{2}}=\frac{p_{2}}{m},\end{eqnarray*}
yielding

\begin{eqnarray}
m\left(\ddot{x}_{1}+\ddot{x}_{2}\right) & =-k\left(x_{1}+x_{2}\right),\label{Eq_mov_1}\end{eqnarray}
and \begin{eqnarray}
m\left(\ddot{x}_{1}-\ddot{x}_{2}\right) & = & -\left(k+2\lambda\right)\left(x_{1}-x_{2}\right).\label{Eq_mov_2}\end{eqnarray}
 The general solutions to (\ref{Eq_mov_1}) and (\ref{Eq_mov_2})
are \[
\sqrt{2}\xi_{+}=x_{1}+x_{2}=A\cos\left(\sqrt{\frac{k}{m}}\textrm{t+}\theta\right),\]
\[
\sqrt{2}\xi_{-}=x_{1}-x_{2}=A^{\prime}\cos\left(\sqrt{\frac{k+2\lambda}{m}}t+\theta^{\prime}\right),\]
($\xi_{+}$ and $\xi_{-}$ are the normal coordinates of the coupled
harmonic oscillators) or, equivalently,

\begin{eqnarray*}
x_{1} & = & \frac{A}{2}\cos\left(\sqrt{\frac{k}{m}}t+\theta\right)+\frac{A^{\prime}}{2}\cos\left(\sqrt{\frac{k+2\lambda}{m}}t+\theta^{\prime}\right)\\
x_{2} & = & \frac{A}{2}\cos\left(\sqrt{\frac{k}{m}}t+\theta\right)-\frac{A^{\prime}}{2}\cos\left(\sqrt{\frac{k+2\lambda}{m}}t+\theta^{\prime}\right).\end{eqnarray*}
 We will assume that the two oscillators are initially at rest the
first one in its equilibrium position (null initial energy, $E_{1}^{i}=0$),
while the second one is dislocated from its equilibrium position by
a distance $D$ (initial energy given by $E_{2}^{i}=\left(1/2\right)kD^{2}$):
\begin{eqnarray*}
\dot{x}_{1}(0)=\dot{x}_{2}(0) & = & 0,\\
x_{1}(0) & = & 0,\\
x_{2}(0) & = & D.\end{eqnarray*}
The integration constants then read

\begin{eqnarray*}
\theta=\theta^{\prime} & = & 0,\\
A & = & D,\\
A^{\prime} & = & -D,\end{eqnarray*}
yielding

\begin{eqnarray}
x_{1}(t) & = & \frac{D}{2}\left[\cos(\omega t)-\cos\left(\omega't\right)\right]\label{Sol_x1}\\
x_{2}(t) & = & \frac{D}{2}\left[\cos\left(\omega t\right)+\cos\left(\omega't\right)\right].\label{Sol_x2}\end{eqnarray}
where we defined $\omega\equiv\sqrt{k/m}$ and $\omega'\equiv\omega\sqrt{1+\varepsilon}$,
with $\varepsilon=2\lambda/k$. Equations (\ref{Sol_x1}) and (\ref{Sol_x2})
can be written in the following suggestive way.\begin{equation}
x_{1}(t)=-D\sin\left[\frac{(\omega-\omega')t}{2}\right]\sin\left[\frac{(\omega+\omega')t}{2}\right],\end{equation}
\begin{equation}
x_{2}(t)=D\cos\left[\frac{(\omega-\omega')t}{2}\right]\cos\left[\frac{(\omega+\omega')t}{2}\right].\end{equation}

We will now assume that the interaction constant $\lambda$ is weak
when compared to the elastic constant $k$, $\varepsilon\ll1$. Then,
we can expand $\omega'$ around $\varepsilon=0,$ yielding\begin{equation}
\omega'=\sqrt{\frac{k+2\lambda}{m}}=\omega\sqrt{1+\varepsilon}=\omega+2\delta\omega\label{delta_omega}\end{equation}
with \begin{equation}
\delta\omega\equiv\frac{\omega'-\omega}{2}\approx\frac{\lambda}{2\sqrt{km}}.\label{eq:delta_omega_2}\end{equation}
Defining \begin{equation}
\bar{\omega}\equiv\frac{\omega'+\omega}{2}=\omega+\delta\omega,\label{eq:omega_bar_2}\end{equation}
the solutions can now be written as

\begin{eqnarray}
x_{1}(t) & = & D\sin(\delta\omega\, t)\sin[\bar{\omega}\, t],\label{Sol_x1_2}\\
x_{2}(t) & = & D\cos(\delta\omega\, t)\cos[\bar{\omega}\, t],\label{Sol_x2_2}\end{eqnarray}
where the dependence on $\lambda$ of Eqs. (\ref{Sol_x1_2}) and (\ref{Sol_x2_2})
are present in $\delta\omega$ and $\bar{\omega}$ through (\ref{eq:delta_omega_2})
and (\ref{eq:omega_bar_2}). 

The movement of both particles is periodic, with two characteristic
frequencies $\delta\omega$ and $\bar{\omega}.$ The frequencies $\delta\omega$
and $\bar{\omega}$ are known as the normal modes of vibration, with
$\bar{\omega}$ being called the higher normal mode and $\delta\omega$
the lower normal mode. Both movements have period $\tau=2\pi/\bar{\omega}$
and are modulated by a variable amplitude with much greater period
given by $\tau=2\pi/\delta\omega$. They are $\pi/2$ out of phase.
We can compute the energy of the two particles, $E_{1}=p_{1}^{2}/2m+kx_{1}^{2}/2$
and $E_{2}=p_{2}^{2}/2m+kx_{2}^{2}/2$. They are \begin{eqnarray}
E_{1}(t) & = & \frac{kD^{2}}{2}\sin^{2}(\delta\omega\, t)\left[1+4\frac{\delta\omega}{\bar{\omega}}\cos^{2}(\bar{\omega}\, t)\right]\label{Classical_energy_x1}\\
E_{2}(t) & = & \frac{kD^{2}}{2}\cos^{2}(\delta\omega\, t)\left[1+4\frac{\delta\omega}{\bar{\omega}}\sin^{2}(\bar{\omega}\, t)\right]\label{Classical_energy_x2}\end{eqnarray}
Due to the coupling, the particles exchange energy between themselves
periodically, with period $\tau=2\pi/\delta\omega$. Each of the oscillators
achieve its minimum energy value when the other have its maximum value.
The maximum value of the energy can be a little bit bigger then $kD^{2}/2$.
This may seem odd, but we must remember that the extra energy is due
to the interaction energy $\lambda(x_{1}-x_{2})^{2}/2=k\varepsilon(x_{1}-x_{2})^{2}/4$.
It is easy to check that if we add this interaction energy to the
sum $E_{1}+E_{2}$ we obtain the total energy of the system \begin{equation}
E_{T}=\frac{kD^{2}}{2}\left(1+2\frac{\delta\omega}{\bar{\omega}}\right)+O(\delta\omega^{2}),\label{Total_energy_classical}\end{equation}
 a value that is constant for the whole movement, as we should expect.
For more details, see Refs.\cite{French,Symon}, where this system
and generalizations of it are analyzed with detail. Of course, as
the Hamiltonian is time independent, energy is always conserved. 

It is also interesting to note that the total energy of the system
depends on the coupling constant, as shown by (\ref{Total_energy_classical}).
A quick analysis of the origin of the {}``extra'' energy shows us
that this happens because of the initial conditions chosen. The initial
conditions from which we obtained (\ref{Total_energy_classical})
have the particle represented by $x_{2}$ off its equilibrium position,
whereas the other particle is at its equilibrium position, with both
particles having zero kinetic energy. This initial condition obviously
imply that the coupling spring, with elastic coefficient $\lambda$,
is also stretched from its equilibrium position, and therefore has
nonzero potential energy at $t=0.$ If we use other initial conditions,
the {}``extra'' energy due to coupling does not appear. For example,
we can choose both particles at an initial position where all spring
have no potential energy (in our case, $x_{1}=x_{2}=0$) and one of
the particles has some kinetic energy while the other particle has
zero kinetic energy. With this set of initial conditions, the energy
transfer from one particle to the other is the same as before, but
no coupling energy is present in the total energy.

\section{Quantum Evolution: Exact Solutions\label{sec:Quantum-Evolution:-Exact}}

Now we want to study the quantized version of the resonant spinless
one-dimensional coupled harmonic oscillator presented in the previous
Section. First we note that the total Hilbert space $\mathcal{H}=\mathcal{H}_{1}\otimes\mathcal{H}_{2}$
is spanned by $\mathcal{H}_{1}$ and $\mathcal{H}_{2}$, the Hilbert
spaces for particles 1 and 2, respectively. For example, the two canonical
variables describing particle $1$ are \[
\hat{X}_{1},\hat{P}_{1}\in\mathcal{H}_{1},\]
 with \[
[\hat{X}_{1},\hat{P}_{1}]=i\hbar\hat{1},\]
 and are therefore represented as \[
\hat{X}_{1}\otimes\hat{1}_{2},\hat{P}_{1}\otimes\hat{1}_{2}\in\mathcal{H},\]
 where $\hat{1}_{2}\in\mathcal{H}_{2}$ is the identity operator.
In this way, the Hamiltonian operator for particle $1$, is written
as \[
\hat{H}_{1}=\frac{1}{2m}\left(\hat{P}_{1}\otimes\hat{1}\right)^{2}+\frac{1}{2}k\left(\hat{X}_{1}\otimes\hat{1}+d\hat{1}\otimes\hat{1}\right)^{2}.\]
 For shortness of notation, we will drop the tensor product and keep
in mind that operators regarding particle 1 act on $\mathcal{H}_{1}$
whereas operators regarding particle 2 act on $\mathcal{H}_{2}$. 

With the simplified notation, the total quantum Hamiltonian operator
for the two oscillators plus the interaction term is \begin{eqnarray}
\hat{H} & = & \hat{H}_{1}+\hat{H}_{2}+\hat{H}_{I}\nonumber \\
 & = & \frac{1}{2m}\hat{P}_{1}^{2}+\frac{1}{2}k\left(\hat{X}_{1}+\hat{d}\right)^{2}+\frac{1}{2m}\hat{P}_{2}^{2}+\frac{1}{2}k\left(\hat{X}_{2}-\hat{d}\right)^{2}+\frac{1}{2}\lambda\left(\hat{X}_{1}-\hat{X}_{2}+2\hat{d}\right)^{2}.\label{Quantum_Hamiltonian}\end{eqnarray}
 We can now make the following change of variables, similar to the
classical case:\begin{eqnarray*}
\hat{x}_{1} & = & \hat{X}_{1}+\hat{d},\\
\hat{x}_{2} & = & \hat{X}_{2}-\hat{d},\\
\hat{p}_{1} & = & \hat{P}_{1},\\
\hat{p}_{2} & = & \hat{P}_{2}.\end{eqnarray*}
This change of variables obviously keeps the commutation relations
between momenta and positions. Hence, in the coordinate representation
we have the Hamiltonian operator \begin{equation}
\hat{H}=-\frac{\hbar^{2}}{2m}\left(\frac{\partial^{2}}{\partial x_{1}^{2}}+\frac{\partial^{2}}{\partial x_{2}^{2}}\right)+\frac{1}{2}k\left(x_{1}^{2}+x_{2}^{2}\right)+\frac{1}{2}\lambda\left(x_{1}-x_{2}\right)^{2}.\end{equation}
 In analogy to the classical case, we work with the normal coordinates
\begin{eqnarray}
\xi_{+} & = & \frac{1}{\sqrt{2}}\left(x_{1}+x_{2}\right),\\
\xi_{-} & = & \frac{1}{\sqrt{2}}\left(x_{1}-x_{2}\right).\end{eqnarray}
 This change of variables has Jacobian one, and does not change the
normalization of wavefunctions. 

With the normal coordinates, the Hamiltonian is \begin{equation}
\hat{H}=-\frac{\hbar^{2}}{2m}\left(\frac{\partial^{2}}{\partial\xi_{+}^{2}}+\frac{\partial^{2}}{\partial\xi_{-}^{2}}\right)+\frac{1}{2}k\xi_{+}^{2}+\frac{1}{2}\left(k+2\lambda\right)\xi_{-}^{2},\end{equation}
and is now separable, i.e., \begin{equation}
\hat{H}=\hat{H}_{+}+\hat{H}_{-},\end{equation}
 where \begin{eqnarray}
\hat{H}_{+} & = & -\frac{\hbar^{2}}{2m}\frac{\partial^{2}}{\partial\xi_{+}^{2}}+\frac{1}{2}k\xi_{+}^{2},\label{QHO1}\\
\hat{H}_{-} & = & -\frac{\hbar^{2}}{2m}\frac{\partial^{2}}{\partial\xi_{-}^{2}}+\frac{1}{2}\left(k+2\lambda\right)\xi_{-}^{2}.\label{QHO2}\end{eqnarray}
 Equations (\ref{QHO1}) and (\ref{QHO2}) are the well known Hamiltonians
for one-dimensional uncoupled harmonic oscillator with frequencies
$\sqrt{k/m}$ and $\sqrt{\left(k+2\lambda\right)/m}$, respectively.

The Schroedinger equation for the system is \begin{equation}
\hat{H}\psi(\xi_{+},\xi_{-},t)=i\hbar\frac{\partial}{\partial t}\psi(\xi_{+},\xi_{-},t).\label{Schroedinger}\end{equation}
 To solve (\ref{Schroedinger})  we need to find its eigenfunctions
and eigenvalues since they form a basis for the Hilbert space. The
general solution can be written as a superposition of the eigenfunctions.
Hence, we need to find the solutions to the time independent Schroedinger
equation \begin{equation}
\hat{H}\psi^{(l)}(\xi_{+},\xi_{-})=\mathcal{E}_{l}\psi^{(l)}(\xi_{+},\xi_{-}),\label{Schroedinger_independent}\end{equation}
where $l$ is an index (perhaps a collective index for both oscillators)
for the energy to be determined. Since $\hat{H}$ is separable, we
can write (\ref{Schroedinger_independent}) as two independent eigenvalue
equations \begin{equation}
\hat{H}_{+}\phi_{+}^{(n)}(\xi_{+})=E_{n}\phi_{+}^{(n)}(\xi_{+})\label{Schroedinger_plus_independent}\end{equation}
and\begin{equation}
\hat{H}_{-}\phi_{-}^{(n')}(\xi_{-})=E'_{n'}\phi_{-}^{(n')}(\xi_{-}),\label{Schroedinger_minus_independent}\end{equation}
where we define \begin{equation}
\psi^{(l)}(\xi_{+},\xi_{-})=\phi_{+}^{(n)}(\xi_{+})\phi_{-}^{(n')}(\xi_{-}),\end{equation}
and \[
\mathcal{E}_{l}=E_{n}+E'_{n'}.\]
 Clearly, $l$ is an index that depends on both $n$ and $n'$, and
for that reason we will write $\psi^{(n,n')}(\xi_{+},\xi_{-})$ instead
of $\psi^{(l)}(\xi_{+},\xi_{-}).$ The eigenfunctions of (\ref{Schroedinger_plus_independent})
and (\ref{Schroedinger_minus_independent}) are well known to be \begin{equation}
\phi_{+}^{(n)}(\xi_{+})=\left(\frac{\sqrt{mk}}{\pi\hbar2^{2n}(n!)^{2}}\right)^{1/4}H_{n}\left[\left(\frac{\sqrt{mk}}{\hbar}\right)^{1/2}\xi_{+}\right]\exp\left[-\frac{\sqrt{mk}\xi_{+}^{2}}{2\hbar}\right],\end{equation}
\begin{eqnarray}
\phi_{-}^{(n')}(\xi_{-}) & = & \left(\frac{\sqrt{m\left(k+2\lambda\right)}}{\pi\hbar2^{2n'}(n'!)^{2}}\right)^{1/4}H_{n'}\left[\left(\frac{\sqrt{m\left(k+2\lambda\right)}}{\hbar}\right)^{1/2}\xi_{-}\right]\exp\left[-\frac{\sqrt{m\left(k+2\lambda\right)}\xi_{-}^{2}}{2\hbar}\right],\end{eqnarray}
and its corresponding eigenvalues are \begin{eqnarray}
E_{n} & = & \hbar\sqrt{\frac{k}{m}}\left(n+\frac{1}{2}\right)\end{eqnarray}
and \begin{eqnarray}
E'_{n'} & = & \hbar\sqrt{\frac{k+2\lambda}{m}}\left(n'+\frac{1}{2}\right),\end{eqnarray}
 where $H_{n}$ are the Hermite polynomials of order $n$ \cite{Bohm_book}. 

The solution to the time dependent Schroedinger equation (\ref{Schroedinger})
is obtained applying the time evolution operator \[
\hat{U}(t,t_{0})=\exp\left(-i\hat{H}(t-t_{0})/\hbar\right)\]
on $\psi(\xi_{+},\xi_{-},t_{0})$. Since $\psi^{(n,n')}(\xi_{+},\xi_{-})=\phi_{+}^{(n)}(\xi_{+})\phi_{-}^{(n')}(\xi_{-})$
form a basis for $\mathcal{H}$, we have \[
\psi(\xi_{+},\xi_{-},t_{0})=\sum_{n,n'=0}^{\infty}C_{n,n'}\psi^{(n,n')}(\xi_{+},\xi_{-}),\]
 \begin{equation}
C_{n,n'}=\int_{-\infty}^{\infty}\int_{-\infty}^{\infty}\phi_{+}^{(n)}(\xi_{+})\phi_{-}^{(n')}(\xi_{-})\psi(\xi_{+},\xi_{-},t_{0})\, d\xi_{+}d\xi_{-},\label{Cnn}\end{equation}
and we used the reality of $\phi_{+}^{(n)}(\xi_{+})\phi_{-}^{(n')}(\xi_{-})$
in the expression for $C_{n,n'}.$ Then, \begin{eqnarray*}
\psi(\xi_{+},\xi_{-},t) & = & \hat{U}(t,t_{0})\psi(\xi_{+},\xi_{-},0)\\
 & = & \sum_{n,n'=0}^{\infty}C_{n,n'}e^{-iE_{n}t/\hbar}\phi_{+}^{(n)}(\xi_{+})e^{-iE'_{n'}t/\hbar}\phi_{-}^{(n')}(\xi_{-})\\
 & = & \sum_{n,n'=0}^{\infty}C_{n,n'}e^{-i\left(E_{n}+E'_{n'}\right)t/\hbar}\psi^{(n,n')}(\xi_{+},\xi_{-}),\end{eqnarray*}
 where $\exp\left(-i\hat{H}t/\hbar\right)=\exp\left(-i\hat{H}_{+}t/\hbar\right)\exp\left(-i\hat{H}_{-}t/\hbar\right)$
since $\left[\hat{H}_{+},\hat{H}_{-}\right]=0$ and we assumed, for
simplicity, that $t_{0}=0$. 

We can now finally go back to the original coordinate system $x_{1}$
and $x_{2}$, and the explicit form for the general solution in this
coordinate system is \begin{eqnarray}
\psi(x_{1},x_{2},t) & = & \sqrt{\frac{m}{\pi\hbar}}\sum_{n,n'=0}^{\infty}C_{n,n'}\left(\frac{\omega}{2^{2n}(n!)^{2}}\right)^{1/4}\left(\frac{\omega'}{2^{2n'}(n'!)^{2}}\right)^{1/4}\times\nonumber \\
 &  & H_{n}\left[\left(\frac{m\omega}{2\hbar}\right)^{1/2}\left(x_{1}+x_{2}\right)\right]H_{n'}\left[\left(\frac{m\omega'}{2\hbar}\right)^{1/2}\left(x_{1}-x_{2}\right)\right]\times\nonumber \\
 &  & \exp\left\{ -\frac{m}{4\hbar}\left[\omega\left(x_{1}+x_{2}\right)^{2}+\omega'\left(x_{1}-x_{2}\right)^{2}\right]\right\} \times\nonumber \\
 &  & \exp\left\{ -i\left[\left(n+\frac{1}{2}\right)\omega+\left(n'+\frac{1}{2}\right)\omega'\right]t\right\} .\label{general_solution}\end{eqnarray}
 where we defined, as before, $\omega=\sqrt{k/m}$ and $\omega'=\sqrt{k+2\lambda/m}$
. The wavefunction (\ref{general_solution}) thus describe spinless
one-dimensional coupled harmonic oscillators with no approximation.

\section{A Simple Example\label{sec:A-Simple-Example}}

We saw in the classical case that two coupled oscillators can transfer
energy to each other. This was clear with the example where at $t=0$
one oscillator had zero mechanical energy while the other one had
nonzero potential energy. As time passes, the mechanical energy of
the former is transferred to the latter. It is interesting to study
the quantum mechanical analogue to this case, i.e., when one quantum
oscillator is in an excited state and the other is in its fundamental
state. 

We will consider as the initial wavefunction the following\begin{equation}
\psi(x_{1},x_{2},0)=\sqrt{\frac{2}{\pi}}\left(\frac{\sqrt{mk}}{\hbar}\right)x_{2}\exp\left[-\frac{\sqrt{mk}\left(x_{1}^{2}+x_{2}^{2}\right)}{2\hbar}\right].\label{initial_state}\end{equation}
The wavefunction (\ref{initial_state}) is an eigenstate of the Hamiltonian\begin{equation}
\hat{H}=\hat{H}_{1}+\hat{H}_{2}\end{equation}
 without the interaction term $\hat{H}_{I}.$ Clearly, $\psi(x_{1},x_{2},0)$
is separable, i.e., it is possible to write $\psi(x_{1},x_{2},0)=\varphi_{1}(x_{1},0)\varphi_{2}(x_{2},0)$.
Since $\hat{H}_{1}$ ($\hat{H}_{2}$) acts only in $\varphi_{1}(x_{1},0)$
($\varphi_{2}(x_{2},0)$), the state $\psi(x_{1},x_{2},0)$ represents
a system where the particle described by $x_{1}$ is in the ground
state and the particle described by $x_{2}$ is in the first excited
state. So, we can think of our example as the following. We have initially
a system of two harmonic oscillators, one in the ground state and
the other in the first excited state. After $t=0$ we suddenly turn
on a interaction between the two oscillators, and as a consequence
we expect to have a {}``transfer of energy'' from one oscillator
to the other, as it happens in the classical case. We will now proceed
to analyze in details this example.

\subsection{Approximate Solution}

To use equation (\ref{general_solution}) we need to find the coefficients
$C_{nn'}$. It is straightforward to compute the coefficients from
(\ref{Cnn}) by just using the orthogonal properties of the Hermite
polynomials and by rewriting (\ref{initial_state}) in the normal
coordinates, yielding \begin{eqnarray}
C_{n,n'} & = & \sqrt{\omega}\left(\frac{\omega}{2^{2n}(n!)^{2}}\right)^{1/4}\left(\frac{\omega'}{2^{2n'}(n'!)^{2}}\right)^{1/4}\sqrt{\frac{2}{\left(\omega+\omega'\right)}}\left(\frac{\omega'-\omega}{\omega+\omega'}\right)^{j}\times\nonumber \\
 &  & \left[\sqrt{\frac{1}{\omega}}\frac{2j!}{j!}\delta_{n',2j}\delta_{1,n}-\sqrt{\frac{2}{\left(\omega+\omega'\right)}}\frac{\left(2j+1\right)!}{j!}\sqrt{\frac{2\omega'}{\omega+\omega'}}\delta_{n',2j+1}\delta_{0,n}\right],\label{coefficients}\end{eqnarray}
 where $\delta_{ij}$ is Kroenecker's delta. 

It is interesting to note that there exists infinite terms of $C_{n,n'}$
that are different from zero. Therefore, if we write down the expression
for the time evolution of the wavefunction after the interaction we
obtain an expression with an infinite number of terms. However, a
close look at the $C_{n,n'}$ coefficients may shed light on how to
deal with this problem. First we see from (\ref{coefficients}) that
only the terms $C_{0,n'}$ and $C_{1,n'}$ are nonzero. If we compute
the ratio between two consecutive nonzero terms, i.e, $C_{0,n'+2}/C_{0,n'}$
and $C_{1,n'+2}/C_{1,n'}$ we obtain\begin{equation}
\frac{C_{0,n'+2}}{C_{0,n'}}=\left(\frac{\omega'-\omega}{\omega+\omega'}\right)\sqrt{\frac{\left(n'+2\right)}{\left(n'+1\right)}},\label{coefficients_0_n}\end{equation}
\begin{equation}
\frac{C_{1,n'+2}}{C_{1,n'}}=\left(\frac{\omega'-\omega}{\omega+\omega'}\right)\sqrt{\frac{\left(n'+1\right)}{\left(n'+2\right)}}.\label{coefficients_1_n}\end{equation}
We note that both ratios (\ref{coefficients_0_n}) and (\ref{coefficients_1_n})
are proportional to $\left(\frac{\omega'-\omega}{\omega+\omega'}\right)$.
Then, if the coupling constant $\lambda$ is small compared to $k$
(weak coupling) we can make an expansion of (\ref{coefficients_0_n})
and (\ref{coefficients_1_n}) around $\lambda=0$ and obtain, up to
first order in $\lambda$, that \[
\left(\frac{\omega'-\omega}{\omega+\omega'}\right)=\frac{\lambda}{2k}+O\left(\lambda^{2}\right).\]
 We conclude that if $\lambda$ is small compared to $k$, as we increase
the value of $n'$, the coefficients $C_{n,n'}$ become less important.
Therefore, it is justifiable to keep only a finite amount of terms
in the expression for $\psi(x_{1},x_{2},t)$ for small $\lambda$.
In our example, we will keep only terms up to first order in $\lambda.$ 

Since we will be working with $\lambda$ small, it is convenient now
to introduce the following parameters already used in the classical
case\[
\delta\omega=\frac{\omega'-\omega}{2},\]
 \[
\bar{\omega}=\frac{\omega'+\omega}{2}.\]
 Then, if $\lambda$ is small, \[
\delta\omega=\frac{\omega\lambda}{2k}+O(\lambda^{2}),\]
 and \[
\frac{\delta\omega}{\bar{\omega}}\ll1.\]
Keeping only terms up to first order in $\frac{\delta\omega}{\bar{\omega}}$,
we have 

\begin{eqnarray}
\psi(x_{1},x_{2},0) & = & \sum_{n,n'=0}^{\infty}C_{n,n'}\psi^{(n,n')}(x_{1},x_{2})\nonumber \\
 & \cong & C_{1,0}\psi^{(1,0)}+C_{0,1}\psi^{(0,1)}+C_{1,2}\psi^{(1,2)}+C_{0,3}\psi^{(0,3)},\label{Wavefunction_approx}\end{eqnarray}
 where \begin{eqnarray}
C_{10} & \cong & \frac{\sqrt{2}}{2},\label{Coefficient_approx_1}\\
C_{01} & \cong & -\frac{\sqrt{2}}{2},\label{Coefficient_approx_2}\\
C_{12} & \cong & \frac{1}{2}\frac{\delta\omega}{\bar{\omega}},\label{Coefficient_approx_3}\\
C_{03} & \cong & -\frac{\sqrt{3}}{2}\frac{\delta\omega}{\bar{\omega}}.\label{Coefficient_approx_4}\end{eqnarray}
We are finally in a position to write, up to first order, the time
dependent wavefunction for the coupled harmonic oscillators. From
(\ref{Wavefunction_approx}) and (\ref{Coefficient_approx_1})--(\ref{Coefficient_approx_4})
it is straightforward to obtain \begin{eqnarray}
\psi(x_{1},x_{2},t) & = & \sqrt{\frac{1}{2\pi}}\frac{m\bar{\omega}}{\hbar}\exp\left\{ -\frac{m\bar{\omega}}{2\hbar}\left[x_{1}^{2}+x_{2}^{2}\right]\right\} \exp\left\{ -i2\bar{\omega}t\right\} \times\nonumber \\
 &  & \left\{ 2i\left(x_{1}+x_{2}\left[\frac{1}{2}-\frac{m\bar{\omega}}{\hbar}x_{1}^{2}\right]\frac{\delta\omega}{\bar{\omega}}\right)\sin\left(\delta\omega t\right)+2\left(x_{2}+x_{1}\left[\frac{1}{2}-\frac{m\bar{\omega}}{\hbar}x_{2}^{2}\right]\frac{\delta\omega}{\bar{\omega}}\right)\cos\left(\delta\omega t\right)\right.\nonumber \\
 &  & +\frac{1}{2}\frac{\delta\omega}{\bar{\omega}}\left(x_{1}+x_{2}\right)\left[\frac{m\bar{\omega}}{\hbar}\left(x_{1}-x_{2}\right)^{2}-1\right]\exp\left\{ -i\left(2\bar{\omega}+\delta\omega\right)t\right\} \nonumber \\
 &  & \left.-\frac{1}{2}\frac{\delta\omega}{\bar{\omega}}\left(x_{1}-x_{2}\right)\left[\frac{m\bar{\omega}}{\hbar}\left(x_{1}-x_{2}\right)^{2}-3\right]\exp\left\{ -i\left(2\bar{\omega}+3\delta\omega\right)t\right\} \right\} +O(\delta\omega^{2}).\label{Wavefunction}\end{eqnarray}
The wavefunction (\ref{Wavefunction}) determines the evolution of
the system. We will now proceed to analyze the system using (\ref{Wavefunction}).

\subsection{Marginal Probabilities}

From (\ref{Wavefunction}) we compute the joint probability density
for $x_{1}$ and $x_{2}$ as a function of $t.$ The joint density
is simply\[
P(x_{1},x_{2},t)=|\Psi(x_{1},x_{2},t)|^{2},\]
and keeping terms up to first order in $\delta\omega$ we have\begin{eqnarray}
P(x_{1},x_{2},t) & = & \frac{1}{2\pi}\left(\frac{m\bar{\omega}}{\hbar}\right)^{2}\exp\left\{ -\frac{m\bar{\omega}}{\hbar}\left[x_{1}^{2}+x_{2}^{2}\right]\right\} \times\nonumber \\
 &  & \left\{ 4\left(x_{2}^{2}+2x_{1}x_{2}\left[\frac{1}{2}-\frac{m\bar{\omega}}{\hbar}x_{2}^{2}\right]\frac{\delta\omega}{\bar{\omega}}\right)\cos^{2}\left(\delta\omega t\right)\right.\nonumber \\
 &  & +4\left(x_{1}^{2}+2x_{1}x_{2}\left[\frac{1}{2}-\frac{m\bar{\omega}}{\hbar}x_{1}^{2}\right]\frac{\delta\omega}{\bar{\omega}}\right)\sin^{2}\left(\delta\omega t\right)\nonumber \\
 &  & +2x_{2}\frac{\delta\omega}{\bar{\omega}}\left(x_{1}+x_{2}\right)\left[\frac{m\bar{\omega}}{\hbar}\left(x_{1}-x_{2}\right)^{2}-1\right]\cos\left(\delta\omega t\right)\cos\left\{ \left(2\bar{\omega}+\delta\omega\right)t\right\} \nonumber \\
 &  & -2x_{2}\frac{\delta\omega}{\bar{\omega}}\left(x_{1}-x_{2}\right)\left[\frac{m\bar{\omega}}{\hbar}\left(x_{1}-x_{2}\right)^{2}-3\right]\cos\left(\delta\omega t\right)\cos\left\{ \left(2\bar{\omega}+3\delta\omega\right)t\right\} \nonumber \\
 &  & -2x_{1}\frac{\delta\omega}{\bar{\omega}}\left(x_{1}+x_{2}\right)\left[\frac{m\bar{\omega}}{\hbar}\left(x_{1}-x_{2}\right)^{2}-1\right]\sin\left(\delta\omega t\right)\sin\left\{ \left(2\bar{\omega}+\delta\omega\right)t\right\} \nonumber \\
 &  & \left.+2x_{1}\frac{\delta\omega}{\bar{\omega}}\left(x_{1}-x_{2}\right)\left[\frac{m\bar{\omega}}{\hbar}\left(x_{1}-x_{2}\right)^{2}-3\right]\sin\left(\delta\omega t\right)\sin\left\{ \left(2\bar{\omega}+3\delta\omega\right)t\right\} \right\} \label{eq:Density_Total}\end{eqnarray}
It is interesting to see how the marginal probability distributions
for $x_{1}$ and $x_{2}$ behave. Let us recall that the marginals
are defined as \begin{equation}
P(x_{1},t)=\int_{-\infty}^{\infty}P(x_{1},x_{2},t)\, dx_{2},\label{Marginal_definition_1}\end{equation}
 and \begin{equation}
P(x_{2},t)=\int_{-\infty}^{\infty}P(x_{1},x_{2},t)\, dx_{1}.\label{Marginal_definition_2}\end{equation}
Therefore, $P(x_{1},t)\, dx_{1}$ represents the probability of measuring
the position of particle 1 in the interval $(x_{1},x_{1}+dx_{1})$
independently of particle 2. The interpretation for $P(x_{2},t)$
is similar. 

From (\ref{eq:Density_Total}), (\ref{Marginal_definition_1}), and
(\ref{Marginal_definition_2}) it is tedious but straightforward to
compute (once again up to first order in $\delta\omega$) such quantities,
which read\begin{eqnarray}
P(x_{1},t) & = & \sqrt{\frac{m\bar{\omega}}{\hbar\pi}}\exp\left\{ -\frac{m\bar{\omega}}{\hbar}x_{1}^{2}\right\} \nonumber \\
 &  & \left\{ \cos^{2}\left(\delta\omega t\right)+\frac{2m\bar{\omega}}{\hbar}x_{1}^{2}\sin^{2}\left(\delta\omega t\right)\right.\nonumber \\
 &  & -\frac{\delta\omega}{\bar{\omega}}\left[\left[\frac{1}{4}-\frac{m\bar{\omega}}{2\hbar}x_{1}^{2}\right]\left(3\cos\left(\left(2\bar{\omega}+3\delta\omega\right)t\right)-\cos\left(\left(2\bar{\omega}+\delta\omega\right)t\right)\right)\cos\left(\delta\omega t\right)\right.\nonumber \\
 &  & \left.\left.-\frac{m\bar{\omega}}{\hbar}x_{1}^{2}\left[\frac{3}{2}-\frac{m\bar{\omega}}{\hbar}x_{1}^{2}\right]\left(\sin\left(\left(2\bar{\omega}+3\delta\omega\right)t\right)-\sin\left(\left(2\bar{\omega}+\delta\omega\right)t\right)\right)\sin\left(\delta\omega t\right)\right]\right\} ,\label{eq:Marginal_x1}\end{eqnarray}
and\begin{eqnarray}
P(x_{2},t) & = & \sqrt{\frac{m\bar{\omega}}{\pi\hbar}}\exp\left\{ -\frac{m\bar{\omega}}{\hbar}x_{2}^{2}\right\} \nonumber \\
 &  & \left\{ \sin^{2}\left(\delta\omega t\right)+\frac{2m\bar{\omega}}{\hbar}x_{2}^{2}\cos^{2}\left(\delta\omega t\right)\right.\nonumber \\
 &  & -\frac{\delta\omega}{\bar{\omega}}\left[\left[\frac{1}{4}-\frac{m\bar{\omega}}{2\hbar}x_{2}^{2}\right]\left(3\sin\left(\left(2\bar{\omega}+3\delta\omega\right)t\right)+\sin\left(\left(2\bar{\omega}+\delta\omega\right)t\right)\right)\sin\left(\delta\omega t\right)\right.\nonumber \\
 &  & \left.\left.-\frac{m\bar{\omega}}{\hbar}x_{2}^{2}\left[\frac{3}{2}-\frac{m\bar{\omega}}{\hbar}x_{2}^{2}\right]\left(\cos\left(\left(2\bar{\omega}+3\delta\omega\right)t\right)+\cos\left(\left(2\bar{\omega}+\delta\omega\right)t\right)\right)\cos\left(\delta\omega t\right)\right]\right\} ,\label{eq:Marginal_x2}\end{eqnarray}
We can compute the values of the marginals (\ref{eq:Marginal_x1})
and (\ref{eq:Marginal_x2}) at $t=0$ and find that, after making
sure that we use $\omega$ as the frequency instead of $\bar{\omega}$,
and keeping only terms up to first order in $\delta\omega/\bar{\omega}$,
such marginals indeed represent the ones for the ground state HO and
the first excited state HO, as one should expect. 

To better grasp the behavior of (\ref{eq:Marginal_x1}) and (\ref{eq:Marginal_x2}),
let us plot them as a function of time. Before plotting, we need to
choose the appropriate values for the constants in the equations.
If our system is in atomic scale, it is not reasonable, from a computational
point of view, to use the MKS system. So, we will measure time in
femtoseconds ($1\,\textrm{fs}=10^{-15}\,\textrm{s}$) and distance
in Angstroms (1 \AA$=10^{-10}\,\textrm{m}$). If we say that the
particles in the oscillators are electrons, then $m=1\,\textrm{m}_{e}$,
where $\textrm{m}_{\textrm{e}}$ is the mass of the electron, then
we have \begin{eqnarray*}
\hbar & = & 10\,\textrm{m}_{e}\cdot{\mbox{\AA}}^{2}\cdot\textrm{fs}^{-1},\end{eqnarray*}
and \[
k=1\,\textrm{m}_{e}\cdot\textrm{fs}^{-2},\]
and, for the harmonic oscillator,%
\begin{figure}[htbp]
\begin{center}\includegraphics[%
  scale=0.3]{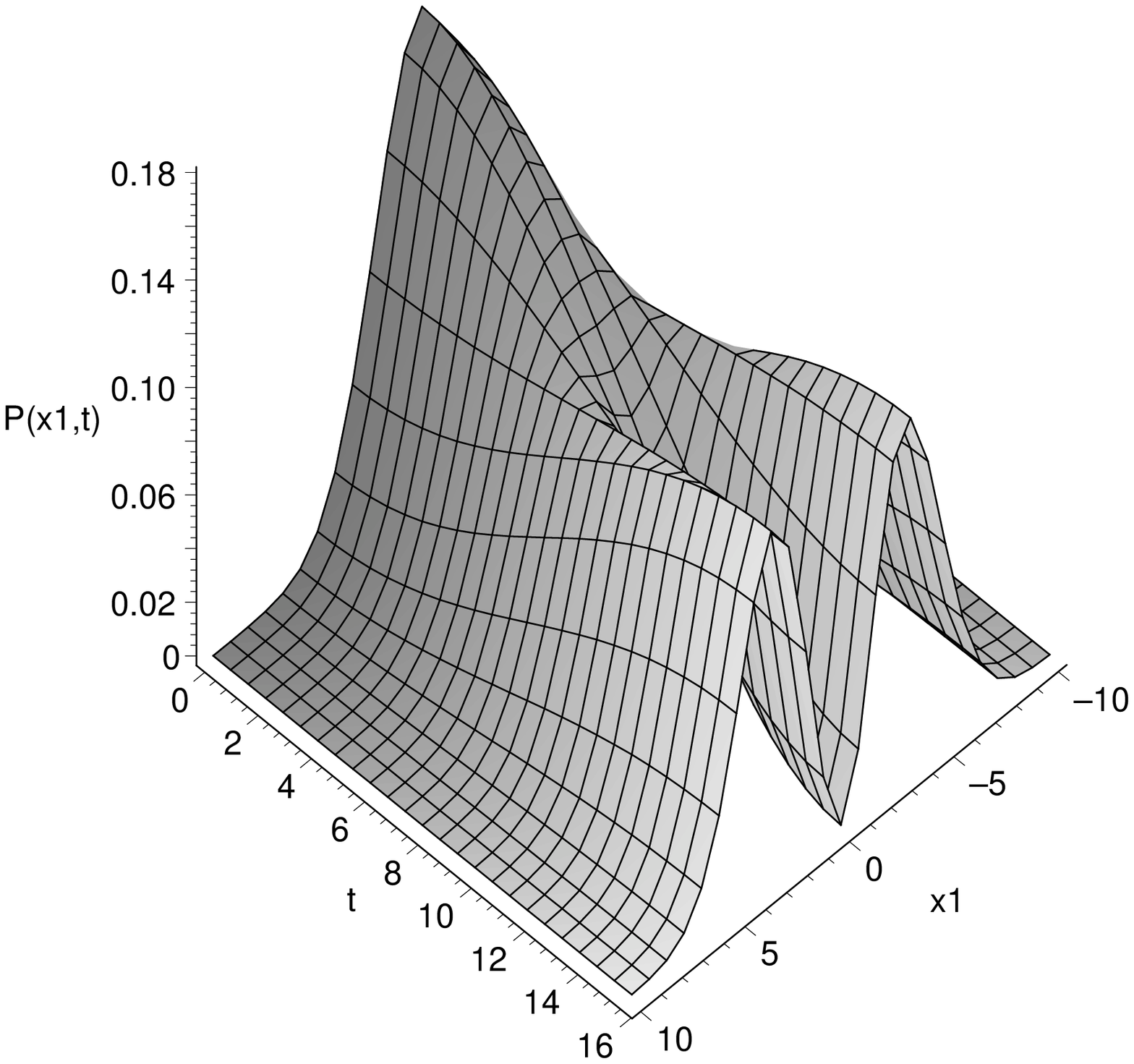}\includegraphics[%
  scale=0.3]{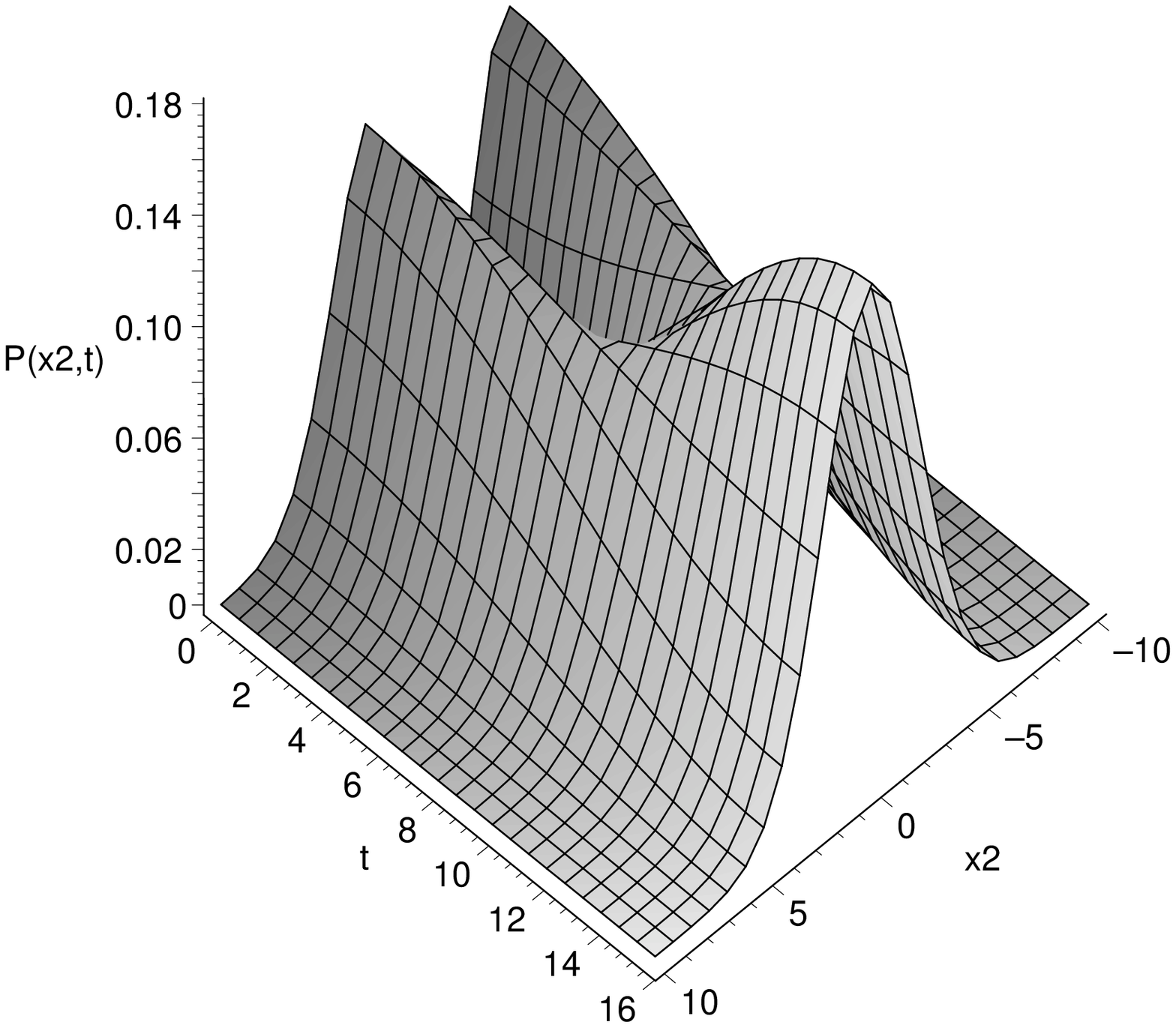}\end{center}

\caption{\label{Fig:Prob_x1}Graphs for the marginal probabilities of $x_{1}$
and $x_{2}$ as a function of time. In these graphs we used $m=1\,\textrm{m}_{\textrm{e}}$,
$\bar{\omega}=1\,\textrm{fs}^{-1}$ and $\delta\omega/\bar{\omega}=1/10$.
The scale for time is fs and the scale for distance is  \AA.}
\end{figure}
 \[
\langle(\Delta x)^{2}\rangle=\frac{\hbar}{2m_{e}\omega}.\]
 The behavior of the probability density for particles $1$ and $2$
are found in Figure \ref{Fig:Prob_x1}. The time interval chosen for
the time axis in the graphs was $\Delta t=\pi/\delta\omega$ as this
is the value where $\cos\left(\delta\omega t\right)=-1$, which is
an extreme in the behavior of the marginal densities. Looking at the
graphs we see that particle $1$ starts with a marginal density that
is mainly a Gaussian function, whereas particle $2$ starts from the
product of $x_{2}^{2}$ times a Gaussian. This is because particle
$1$ is at the ground state and particle $2$ is at the first excited
state at $t=0.$ However, as time passes there is a swap in the roles
of particle $1$ and $2,$ in the sense that at $t=\pi/\delta\omega$
the marginal density for particle $1$ resembles that of particle
$2$ for $t=0$ and vice versa. This is of course due to the interaction
between the two particles. We may think of those densities as showing
that, at $t=\pi/\delta\omega$ (more generally when $t=\left(2n+1\right)\pi/\delta\omega$)
particle $1$ is no longer in the ground state, but in the first excited
state, whereas particle $2$ is in the ground state.

\subsection{Energy Expectations}

The densities above suggest that there is an energy transfer from
one particle to the other. To see that this is the case, let us compute
the energy values for each particle. First we should note that the
system is not in an eigenstate of the Hamiltonian, as we started from
a superposition of different energy states. We define the energy or
particle $1$ as \[
E_{1}=\langle\hat{H}_{1}\rangle,\]
the energy of particle $2$ as \[
E_{2}=\langle\hat{H}_{2}\rangle,\]
and the total energy as the sum of the two energies plus the interaction
energy\[
E_{T}=E_{1}+E_{2}+\langle\hat{H}_{I}\rangle.\]
 In coordinate representation we have that\begin{eqnarray*}
E_{1} & = & \int_{-\infty}^{\infty}\int_{-\infty}^{\infty}dx_{1}dx_{2}\,\psi(x_{1},x_{2},t)^{*}\hat{H}_{1}\psi(x_{1},x_{2},t)\\
 & = & \int_{-\infty}^{\infty}\int_{-\infty}^{\infty}dx_{1}dx_{2}\,\psi(x_{1},x_{2},t)^{*}\left[-\frac{\hbar^{2}}{2m}\frac{\partial^{2}}{\partial x_{1}^{2}}+\frac{1}{2}kx_{1}^{2}\right]\psi(x_{1},x_{2},t),\end{eqnarray*}
 and computing this term we obtain, up to second order in $\delta\omega/\bar{\omega}$,
\begin{eqnarray}
E_{1} & = & \hbar\bar{\omega}\left(\frac{1}{2}+\sin^{2}\left(\delta\omega t\right)\right)\left(1-\delta\omega/\bar{\omega}\right)\nonumber \\
 & = & \hbar\omega\left(\frac{1}{2}+\sin^{2}\left(\delta\omega t\right)\right).\label{eq:Quantum_energy_x1}\end{eqnarray}
Similarly, for $E_{2}$ we have \begin{eqnarray}
E_{2} & = & \hbar\bar{\omega}\left(\frac{1}{2}+\cos^{2}\left(\delta\omega t\right)\right)\left(1-\delta\omega/\bar{\omega}\right).\nonumber \\
 & = & \hbar\omega\left(\frac{1}{2}+\cos^{2}\left(\delta\omega t\right)\right).\label{eq:Quantum_energy_x2}\end{eqnarray}
 If we compare the quantum energies (\ref{eq:Quantum_energy_x1})
and (\ref{eq:Quantum_energy_x2}) to the classical expressions (\ref{Classical_energy_x1})
and (\ref{Classical_energy_x2}) the resemblance is striking. They
are practically the same for $\delta\omega/\bar{\omega}\ll1$, except
for a zero energy factor of $\frac{1}{2}\hbar\bar{\omega}$ present
in the quantum mechanical case. In fact, the same conclusions can
now be drawn from (\ref{eq:Quantum_energy_x1}) and (\ref{eq:Quantum_energy_x2})
, i.e., that due to the coupling, the particles exchange energy between
themselves periodically, with period $\tau=2\pi/\delta\omega$. Each
of the oscillators achieve its minimum energy value when the other
have its maximum value. For the interaction energy we compute \begin{equation}
\langle\hat{H}_{I}\rangle=2\hbar\delta\omega.\label{Quantum_interaction_term}\end{equation}
 Then, it is easy to compute the total mean energy\begin{eqnarray*}
E_{T} & = & E_{1}+E_{2}+\langle\hat{H}_{I}\rangle\\
 & = & 2\hbar\bar{\omega}\\
 & = & 2\hbar\omega+2\hbar\delta\omega.\end{eqnarray*}
This is once again in agreement with the classical case seen above,
in the sense that the total energy is the sum of the energy of each
oscillator (keeping into account the nonclassical zero point energy)
without the interaction term plus an interaction term $2\hbar\delta\omega$. 

We just saw that the state we used had a term in the total energy
$2\hbar\delta\omega$ that was due to the coupling between the two
oscillators. However, if we remember the classical case of Section
2, with different initial conditions --- e.g. $x_{1}=0,$ $x_{2}=0$,
$\dot{x}_{1}=v$, $\dot{x}_{2}=0$, at $t=0$ --- no interaction term
is present in the total energy. What about the quantum case? Do we
always have an interaction term present, as in (\ref{Quantum_interaction_term})?
A short computation shows that for any initial state that is a combination
of Fock states for the two HO of the form \[
|\psi\rangle=|n_{1}\rangle\otimes|n_{2}\rangle,\]
 where $|n_{1}\rangle$ and $|n_{2}\rangle$ are eigenstates of two
uncoupled HO, the value of $\langle\hat{H}_{I}\rangle_{\psi}$ (the
interaction term) is different from zero. 

The question remains as to whether it is possible to find an initial
state that has an interaction term that is zero. A good guess would
be to take both HO in a coherent state at $t=0$, since it is a state
that has many of the characteristics of a classical system \cite{Cohen_QM}.
It is easy to show that it is indeed true that for the state\[
|\psi\rangle=|\alpha\rangle\otimes|\beta\rangle,\]
where \[
|\alpha\rangle=e^{-\frac{|\alpha|^{2}}{2}}\,\sum_{n=0}^{\infty}\frac{\alpha^{n}}{\sqrt{n!}}|n\rangle,\]
 and similar for $|\beta\rangle$, the expected value of the interaction
energy at $t=0$ is zero if $\alpha$ and $\beta$ have an appropriate
phase relation. It is left up to the reader to find out this phase
relation and a set of initial conditions for a classical system which
reproduces the expectations in the quantum mechanical case.

\section{The Bohmian Interpretation\label{sec:Bohm}}

Before we analyze the transfer of energy from a Bohmian point of view,
let us quickly review Bohm's interpretation of quantum mechanics.
Let us begin with the causal interpretation for the case of the Schr\"{o}dinger
equation describing a single particle. In the coordinate representation,
for a non-relativistic particle with Hamiltonian $\hat{H}=\hat{p}^{2}/2m+V(\hat{x}),$
the Schr\"{o}dinger equation is

\begin{equation}
i\hbar\frac{\partial\Psi(x,t)}{\partial t}=\left[-\frac{\hbar^{2}}{2m}\nabla^{2}+V(x)\right]\Psi(x,t).\label{bsc}\end{equation}
 We can transform this differential equation over a complex field
into a pair of coupled differential equations over real fields. We
do that by writing $\Psi=R\exp(iS/\hbar)$, where $R$ and $S$ are
real functions, and substituting it into (\ref{bsc}). We obtain the
following equations. \begin{equation}
\frac{\partial S}{\partial t}+\frac{(\nabla S)^{2}}{2m}+V-\frac{\hbar^{2}}{2m}\frac{\nabla^{2}R}{R}=0,\label{bqp}\end{equation}
\begin{equation}
\frac{\partial R^{2}}{\partial t}+\nabla\cdot(R^{2}\frac{\nabla S}{m})=0.\label{bpr}\end{equation}
 The usual probabilistic interpretation, i.e. the Copenhagen interpretation,
understands equation (\ref{bpr}) as a continuity equation for the
probability density $R^{2}$ for finding the particle at position
$x$ and time $t$. All physical information about the system is contained
in $R^{2}$, and the total phase $S$ of the wave function is completely
irrelevant. In this interpretation, nothing is said about $S$ and
its evolution equation (\ref{bqp}). 

However, examining equation (\ref{bpr}), we can see that $\nabla S/m$
may be interpreted as a velocity field, suggesting the identification
$p=\nabla S$. Hence, we can look to equation (\ref{bqp}) as a Hamilton-Jacobi
equation for the particle with the extra potential term \[
Q=-\frac{\hbar^{2}}{2m}\frac{\nabla^{2}R}{R},\]
where $Q$ is the so called quantum potential. Thus, since Bohm's
interpretation identifies $p$ with $\nabla S$, from the differential
equation $p=m\dot{x}=\nabla S$ we may compute its solutions and obtain
the trajectory of the quantum particle. Therefore, in Bohm's interpretation
both momentum and position are quantities that are ontologically well
defined. 

For our case of two coupled-HO, the configuration space has two variables,
$x_{1}$ and $x_{2}$, representing the positions of particles 1 and
2, respectively. For two particles, the nonlocality of Bohm's interpretation
becomes evident as the Schr\"{o}dinger equation becomes

\begin{equation}
i\hbar\frac{\partial\Psi(x_{1},x_{2},t)}{\partial t}=\left[-\frac{\hbar^{2}}{2m_{1}}\nabla_{1}^{2}-\frac{\hbar^{2}}{2m_{2}}\nabla_{2}^{2}+V(x_{1},x_{2})\right]\Psi(x_{1},x_{2},t),\end{equation}
 where $\nabla_{i}^{2}$ is the laplacian operator with respect to
the coordinates of particle $i.$ If we follow the same transformation
as before, we can obtain the following equations. \begin{equation}
\frac{\partial S}{\partial t}+\frac{(\nabla_{1}S)^{2}}{2m_{1}}+\frac{(\nabla_{2}S)^{2}}{2m_{2}}+V-\frac{\hbar^{2}}{2m}\frac{\nabla^{2}R}{R}=0,\end{equation}
\begin{equation}
\frac{\partial R^{2}}{\partial t}+\nabla_{1}\cdot\left(R^{2}\frac{\nabla_{1}S}{m_{1}}\right)+\nabla_{2}\cdot\left(R^{2}\frac{\nabla_{2}S}{m_{2}}\right)=0.\end{equation}
The nonlocality comes from the fact that, even if the potential $V(x_{1},x_{2})$
is local, it is possible that the quantum potential given by \[
Q=-\frac{\hbar^{2}}{2m_{1}}\frac{\nabla_{1}^{2}R}{R}-\frac{\hbar^{2}}{2m_{2}}\frac{\nabla_{2}^{2}R}{R}\]
is nonlocal, depending on the form of $R.$ This characteristic is
necessary, as proved by Bell, if Bohm's theory is to recover all quantum
mechanical predictions. 

Using (\ref{Wavefunction}) it is straightforward to compute the phase
$S(x_{1},x_{2},t)$ from the expression\[
S(x_{1},x_{2},t)=-\hbar\,\arctan\left[-i\,\frac{\Psi(x_{1},x_{2},t)-\Psi(x_{1},x_{2},t)^{*}}{\Psi(x_{1},x_{2},t)+\Psi(x_{1},x_{2},t)^{*}}\right].\]
After some long and tedious algebra we obtain\[
S(x_{1},x_{2},t)=-\hbar\,\arctan\left(\frac{S_{A}(x_{1},x_{2}t)}{S_{B}(x_{1},x_{2}t)}\right),\]
where \begin{eqnarray*}
S_{A}(x_{1},x_{2},t) & = & 4\cos\left(2\bar{\omega}t\right)\left\{ \left(x_{1}^{2}\sin\left(\delta\omega t\right)^{2}-x_{2}^{2}\cos\left(\delta\omega t\right)^{2}\right)\sin\left(2\bar{\omega}t\right)\right.\\
 &  & \left.+x_{1}x_{2}\sin\left(\delta\omega t\right)\cos\left(\delta\omega t\right)\right\} ,\end{eqnarray*}
and 

\begin{eqnarray*}
S_{B}(x_{1},x_{2},t) & = & \left(x_{2}\cos(2\omega t)\cos(\delta\omega t)+x_{1}\sin(2\omega t)\sin(\delta\omega t)\right)^{2},\end{eqnarray*}
where we keept all terms in $\left(\delta\omega/\bar{\omega}\right)t$
but we neglected terms in $\delta\omega/\bar{\omega}$. 

\begin{figure*}[htbp]
\begin{center}\includegraphics[%
  scale=0.5]{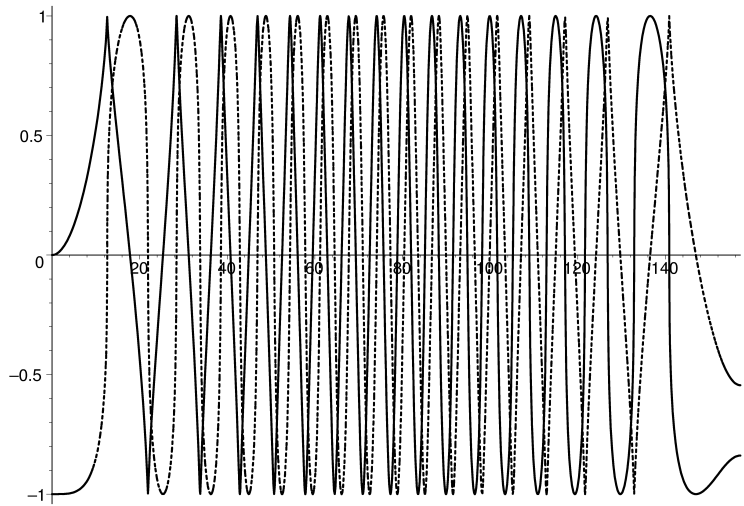}\end{center}

\caption{Bohmian trajectories for two CHO. The trajectories correspond to
$\bar{\omega}=1\,\textrm{fs}^{-1}$, $\delta\omega/\bar{\omega}=0.01$,
$x_{1}(0)=0$, and $x_{2}(0)=-1$. The solid line represents the trajectory
of $x_{1}(t)$ whereas the dashed line represents that of $x_{2}(t)$.
The scale for the ordinates is in $\mbox{\AA}$ and the time scale
is in fs. \label{cap:Bohmian-trajectories-delta1}}
\end{figure*}

\begin{figure*}[htbp]
\begin{center}\includegraphics[%
  scale=0.5]{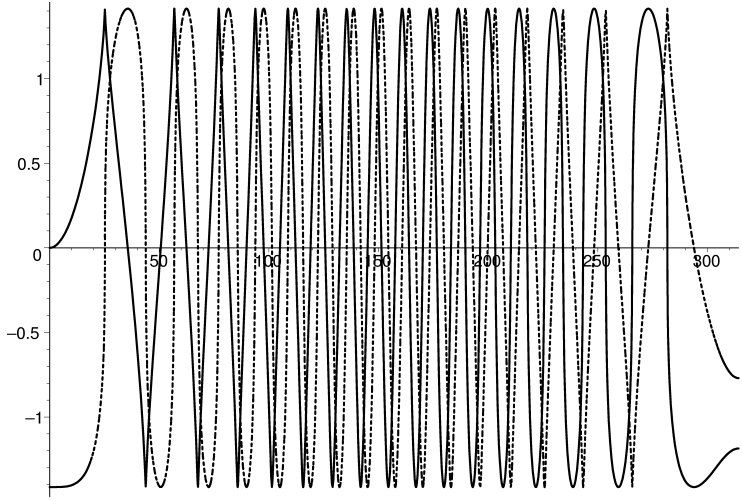}\end{center}

\caption{Bohmian trajectories for two CHO. The trajectories correspond to
$\bar{\omega}=1\,\textrm{fs}^{-1}$, $\delta\omega/\bar{\omega}=0.005$,
$x_{1}(0)=0$, and $x_{2}(0)=-\sqrt{2}$. The solid line represents
the trajectory of $x_{1}(t)$ whereas the dashed line represents that
of $x_{2}(t)$. The scale for the ordinates is in $\mbox{\AA}$ and
the time scale is in fs. We can observe that the trajectories are
identical to the ones shown in the previous Figure, except for the
coordinate scales, a result consistent with equations (\ref{eq:transformation_t})--(\ref{eq:transformation_x2}).
\label{cap:Bohmian-trajectories-delta2} }
\end{figure*}

From $S(x_{1},x_{2}t)$ we obtain the differential equation that describes
the trajectories of particles $x_{1}$ and $x_{2}$ as\begin{equation}
\frac{\textrm{d}x_{1}}{\textrm{d}t}=\frac{1}{m}\frac{\partial S(x_{1},x_{2}t)}{\partial x_{1}}=-\frac{\hbar}{m}\frac{x_{2}\cos\left(\delta\omega t\right)\sin\left(\delta\omega t\right)}{x_{1}^{2}\sin^{2}\left(\delta\omega t\right)+x_{2}^{2}\cos^{2}\left(\delta\omega t\right)}\label{eq:Bohm_differential_eq_1}\end{equation}
and\begin{equation}
\frac{\textrm{d}x_{2}}{\textrm{d}t}=\frac{1}{m}\frac{\partial S(x_{1},x_{2}t)}{\partial x_{2}}=\frac{\hbar}{m}\frac{x_{1}\cos\left(\delta\omega t\right)\sin\left(\delta\omega t\right)}{x_{1}^{2}\sin^{2}\left(\delta\omega t\right)+x_{2}^{2}\cos^{2}\left(\delta\omega t\right)}.\label{eq:Bohm_differential_eq_2}\end{equation}
 We can see that the trajectories follow a set of differential equations
that are coupled and nonlinear. It is interesting to notice that if
$\delta\omega=0$ we recover the standard Bohmian result that in the
case of no interaction each HO is in an eigenstate and therefore both
particles are at rest. However, if $\delta\omega\neq0$, we obtain
at once that, after the change of variables \begin{eqnarray}
t' & = & \frac{\delta\omega}{\delta\omega'}t,\label{eq:transformation_t}\\
x_{1}' & = & \sqrt{\frac{\delta\omega}{\delta\omega'}}x_{1},\\
x_{2}' & = & \sqrt{\frac{\delta\omega}{\delta\omega'}}x_{2},\label{eq:transformation_x2}\end{eqnarray}
 the differential equations (\ref{eq:Bohm_differential_eq_1}) and
(\ref{eq:Bohm_differential_eq_2}) are form invariant with respect
to a change in the coupling constant from $\delta\omega$ to $\delta\omega'$.
This invariance is illustrated in Figures \ref{cap:Bohmian-trajectories-delta1}
and \ref{cap:Bohmian-trajectories-delta2}, where typical Bohmian
trajectories were computed for both particles. The solutions shown
in Figures \ref{cap:Bohmian-trajectories-delta1} and \ref{cap:Bohmian-trajectories-delta2}
were obtained numerically using a 7th-8th-order continuous Runge-Kuta
method. %

It is important to compute, in Bohmian theory, the quantum potential
$Q$ defined as\begin{eqnarray*}
Q & = & Q_{1}+Q_{2}\end{eqnarray*}
where \[
Q_{1}=-\frac{\hbar^{2}}{2m}\frac{1}{\sqrt{P(x_{1},x_{2},t)}}\frac{\partial^{2}\sqrt{P(x_{1},x_{2},t)}}{\partial x_{1}^{2}}\]
and\[
Q_{2}=-\frac{\hbar^{2}}{2m}\frac{1}{\sqrt{P(x_{1},x_{2},t)}}\frac{\partial^{2}\sqrt{P(x_{1},x_{2},t)}}{\partial x_{2}^{2}}.\]
 It is straightforward to compute \begin{eqnarray}
Q_{1} & = & \hbar\bar{\omega}-\frac{1}{2}m\bar{\omega}^{2}x_{1}^{2}\nonumber \\
 &  & +\frac{1}{2}\hbar\bar{\omega}\frac{x_{1}^{2}\sin^{2}\left(\delta\omega t\right)-x_{2}^{2}\cos^{2}\left(\delta\omega t\right)}{x_{1}^{2}\sin^{2}\left(\delta\omega t\right)+x_{2}^{2}\cos^{2}\left(\delta\omega t\right)}\nonumber \\
 &  & -\frac{1}{2}\frac{\hbar^{2}}{m}\frac{x_{2}^{2}\cos^{2}\left(\delta\omega t\right)\sin^{2}\left(\delta\omega t\right)}{\left(x_{1}^{2}\sin^{2}\left(\delta\omega t\right)+x_{2}^{2}\cos^{2}\left(\delta\omega t\right)\right)^{2}},\label{eq:Q1}\end{eqnarray}
and\begin{eqnarray}
Q_{2} & = & \hbar\bar{\omega}-\frac{1}{2}m\bar{\omega}^{2}x_{2}^{2}\nonumber \\
 &  & +\frac{1}{2}\hbar\bar{\omega}\frac{x_{2}^{2}\cos^{2}\left(\delta\omega t\right)-x_{1}^{2}\sin^{2}\left(\delta\omega t\right)}{x_{2}^{2}\cos^{2}\left(\delta\omega t\right)+x_{1}^{2}\sin^{2}\left(\delta\omega t\right)}\nonumber \\
 &  & -\frac{1}{2}\frac{\hbar^{2}}{m}\frac{x_{1}^{2}\sin^{2}\left(\delta\omega t\right)\cos^{2}\left(\delta\omega t\right)}{\left(x_{2}^{2}\cos^{2}\left(\delta\omega t\right)+x_{1}^{2}\sin^{2}\left(\delta\omega t\right)\right)^{2}},\label{eq:Q2}\end{eqnarray}
 which yields \begin{eqnarray}
Q(x_{1},x_{2},t) & = & 2\hbar\bar{\omega}-\frac{1}{2}m\bar{\omega}^{2}\left(x_{1}^{2}+x_{2}^{2}\right)\nonumber \\
 &  & -\frac{1}{2}\frac{\hbar^{2}}{m}\frac{\left(x_{1}^{2}+x_{2}^{2}\right)\sin^{2}\left(\delta\omega t\right)\cos^{2}\left(\delta\omega t\right)}{\left(x_{2}^{2}\cos^{2}\left(\delta\omega t\right)+x_{1}^{2}\sin^{2}\left(\delta\omega t\right)\right)^{2}}.\label{eq:Q}\end{eqnarray}
We are now in a position to compute the total bohmian energy for each
one of the particles,

\begin{eqnarray*}
E_{1} & = & K_{1}+V_{1}+Q_{1}\\
 & = & \hbar\bar{\omega}+\frac{1}{2}\hbar\bar{\omega}\frac{x_{1}^{2}\sin^{2}\left(\delta\omega t\right)-x_{2}^{2}\cos^{2}\left(\delta\omega t\right)}{x_{1}^{2}\sin^{2}\left(\delta\omega t\right)+x_{2}^{2}\cos^{2}\left(\delta\omega t\right)},\end{eqnarray*}
\begin{eqnarray*}
E_{2} & = & K_{2}+V_{2}+Q_{2}\\
 & = & \hbar\bar{\omega}-\frac{1}{2}\hbar\bar{\omega}\frac{x_{1}^{2}\sin^{2}\left(\delta\omega t\right)-x_{2}^{2}\cos^{2}\left(\delta\omega t\right)}{x_{1}^{2}\sin^{2}\left(\delta\omega t\right)+x_{2}^{2}\cos^{2}\left(\delta\omega t\right)},\end{eqnarray*}
where $K_{i}=\frac{1}{2}m\left(\frac{\textrm{d}x_{i}}{\textrm{d}t}\right)^{2}$
is the kinetic energy of particle $i$ (obtained from the guidance
equations (\ref{eq:Bohm_differential_eq_1}) and (\ref{eq:Bohm_differential_eq_2}))
and $V_{i}$ is the potential for particle $i$ (neglecting terms
in $\delta\omega/\bar{\omega}$). 

The total energy for the system is just the sum of the individiual
energies, yielding\[
E_{T}=E_{1}+E_{2}=2\hbar\bar{\omega},\]
the same value as the expected energy of the system.

\section{Conclusions and Final Remarks\label{sec:Conclusions}}

\textcolor{black}{We see that the expressions obtained for $E_{1}$
and $E_{2}$ involve an interaction term that makes it impossible
to distinguish what part of the energy belongs to the particle $x_{1}$
and what part belongs to the particle $x_{2}$, except for some particular
values of $t$. In the Copenhagen interpretation of QM it does not
make any sense to talk about the energy of each oscillator for all
$t$, as the oscillators are in a quantum superposition and are not
in an eigenstate of its hamiltonian operator. In Bohm, it will not
make any sense to talk about the energy of each oscillator for all
$t,$ since the quantum potential creates an interaction between the
two oscillators that is of the same order of the other terms in the
hamiltonian. Therefore it does not make any sense in the bohmian theory
to say that the energy of the photon was transfered to the photodetector
(except for very special values of $t$).}

However, the bohmian interpretation gives an onthological explanation
for the indefiniteness of the energy of each particle. Even with the
interaction turned off, there is still a quantum nonlocal interaction
between the oscillators given by the quantum potential and, in fact,
one oscillator is not isolated from the other. This indicates that
a real measurement has not yet ocurred. It seems to us that in order
for a measurement to take place, a more elaborated description of
the photodetection process involving a thermal bath or a macroscopic
description must be used. In such case, we expect that the quantum
potential will vanish and no further nonlocal interaction will be
present after the measurement.

\end{document}